\documentstyle[preprint,aps]{revtex}

\begin{document}
\title{Midinfrared Conductivity in  Orientationally  \\
Disordered Doped Fullerides
}
\author{M. S. Deshpande and  E. J. Mele}
\address{ Department of Physics \\
Laboratory for Research on the Structure of Matter\\
University of Pennsylvania,
Philadelphia, PA 19104}
\author{ M. J. Rice}
\address{ Xerox Webster Research Center \\
Webster, New York 14580 }
\author{ H-Y Choi}
\address{ Department of Physics  \\
Sung Kyun Kwan University \\ Suwon 440 746 Korea }

\maketitle
\begin{abstract}
The coupling
between the intramolecular vibrational modes and the doped
conduction
electrons in $M_3C_{60}$ is studied by a calculation of the
electronic contributions to
the phonon self energies.  The calculations are carried out
 for an orientationally
ordered reference solid with symmetry $Fm \bar{3} m$ and for a model with
quenched orientational disorder on the fullerene sites.
In both cases, the dispersion and
symmetry of the renormalized modes is governed by the electronic contributions.
The
current current correlation functions and frequency dependent
conductivity through the
midinfrared are calculated for both models.  In the disordered structures, the
renormalized modes derived from even parity intramolecular phonons are resonant
with the dipole excited single particle spectrum, and modulate the predicted
midinfrared
conductivity. The spectra for this coupled system are calculated
for several recently
proposed microscopic models for the electron phonon coupling,
and a comparison is
made with recent experimental data which demonstrate this effect.

\vspace{0.2in}

PACS Numbers : 78.20.-e, 74.70 Ya, 74.20.-z
\end{abstract}

\vspace{0.5in}


\section{Introduction}

	Soon after the discovery of solid phases of $C_{60}$
 \cite{c60Dis} it was demonstrated
that these solids can be doped from an insulating parent phase through a
series of interesting conducting and nonconducting phases by the
introduction of alkali species in the interstitial volume \cite{MxC60Dis}.
 There has been
particular interest in the properties of the family of $M_3C_{60}$ (where
M is an alkali metal) which exhibit superconducting ground states. The
interplay of the superconductivity and the unique structural properties
exhibited by these molecular metals is a key issue for these materials.

	Among the various proposed theories for the superconductivity of these
phases is a model which posits that an effective attractive interaction
between the conduction electrons is mediated by exchange of
intramolecular phonons in the doped solid \cite{Varma,Schluter1}.
Since this model was originally
proposed, it has been appreciated that the theory can be tested
quantitatively by examining the renormalization of the intramolecular
vibrational modes which derive from their coupling to the doped
conduction
electrons.  Indeed, the key Raman active vibrational modes are
experimentally found to shift and broaden in the conducting state
\cite{Raman1,Raman2,Raman3},
confirming the underlying idea that coupling between the intramolecular
modes and the conduction charge is important for the physics of the doped
phases.

	This theoretical picture is complicated in an essential way by the
orientational ``merohedral" disorder \cite{Disordexp} of the alkali doped
samples which
have been synthesized. Experimentally, the structure of most of the
$M_3C_{60}$ samples appears to be that of an orientational glass.  In
these solids, each of the $C_{60}$ molecules, on the lattice sites of an
FCC
Bravais lattice, adopts one of tho high symmetry orientations with the
two fold molecular axes oriented along the (001) crystal directions. There
are two inequivalent ways in which each molecule can achieve this
orientation, and they are related by a $\pi /2$ rotation about any
 (001) direction. This
choice of orientation varies, essentially randomly, from site to site in
the solid.

	This structural disorder has important consequences for virtually all
the electronic properties of the doped solid, including the renormalization
of the intramolecular modes mentioned above.  Nevertheless, the effects of the
disorder are awkward to treat properly in the theory, and consequently
most theoretical work has treated the effects of orientational disorder only
very approximately. Indeed, a considerable body of work has investigated the
``zero band width" limit of this problem \cite{Varma,Schluter1,Schluter2},
in which the intermolecular
hopping amplitudes are set to zero and progress can be made
analytically.  This approximation is based on the fact that the
intermolecular
electronic hopping amplitudes are much smaller than the intramolecular
amplitudes which allows a useful separation of energy scales.  Thus many
solid state properties do reflect the underlying molecular structure of
the solid.  For
example, the conduction band relevant to our discussion below is derived
primarily from the antibonding $t_{1u}$ molecular state \cite{t1u},
 with very little
mixing between different intramolecular orbitals in the solid phase.
Similarly, each
of the vibrational bands in the solid may be identified with a
fundamental
vibrational excitation of the isolated $C_{60}$.  This intramolecular
symmetry also strongly constrains the possible couplings between the
vibrational degrees of freedom and the conduction electrons
: electronic density fluctuations on a given fullerene site may
be coupled only to the fully symmetric $a_g$ or quadrupolar $h_g$
vibrational excitations on the same site \cite{Varma}.

	Of course the zero band width limit does not address any phenomenon
which derives from the intermolecular properties of the doped solid.  For
example,
the two point density density response function for the doped carriers
depends explicitly on the three dimensional dynamics of the doped
carriers
in these molecular solids.  This in turn depends sensitively on the
degree and type of microscopic structural disorder in the doped phases,
so that spectroscopy on the low energy electronic excitations can
be used  to probe
the structural disorder in the doped solids \cite{GelLu1,GelLu2}. In this
theory, scattering from
static orientational fluctuations in the doped solid  leads to a residual
mean
free path for the conduction charge of order 20 A \cite{MeleErwin1},
and to a residual
resistivity of order 3000 S/cm \cite{GelLu2}.
This strong scattering is believed to
dominate the observed low frequency conductivity which is still lower, of
order 1500 S/cm \cite{Degiorgi,ResisT} at low temperatures.

	The phonon renormalizations mentioned previously are  obtained from
a similar family of two point functions, and are also quite sensitive to
the
intersite orientational correlations of the solid.   For a reference
orientationally ordered solid, the high symmetry of the structure leads
to
selection rules which are applicable to the solid phase and which further
restrict the coupling of the vibrational and electronic degrees of
freedom.
These selection rules can be violated in the disordered solid, depending
on
the type and degree of the orientational disorder. This is very
important for
understanding the renormalization of the intramolecular vibrations in
both the
ordered and disordered phases, and the response functions derived from
these modes \cite{RCDM}.  In fact, the spectrososcopic signatures
of this effect are now
beginning to be reported in experiments for the doped fullerides
where several of the
even parity ``on ball" vibrations can be observed as modulations in the
dipole
active long wavelength and low frequency electronic excitation spectra.
This is strictly forbidden for the orientationally ordered structure, and
its
appearance in experiments provides a powerful  direct probe of the
scattering by oreintational fluctuations in the doped phases.   This
effect is
reminscent of similar phenomena which have been previously encountered
in the charge transfer salts\cite{Rice,MJRice,RPB}, and in the
doped conducting polymers \cite{salts},
although the structural origins of the symmetry breaking here are
somewhat unique to the fullerides.

	In this paper, we present a study of the electron phonon
coupling in the
doped solids in which we explicitly treat the three dimensional dynamics
of the doped conduction charge.  This theory is carried out first for a
reference
orientationally ordered structure, and then for a model with
orientational
randomness.  Below we calculate the electronic contributions to the
phonon
self energies in both models, so that the effects on orientational
disorder on
the phonon self energies can be clearly identified.  Finally, we compute
the
phonon contributions to the long wavelength current current response
function, and study the infrared conductivity for this coupled electron
phonon system.  A brief comparison with experiments is given at the end
of
the paper.   There we find that the mid-infrared conductivity
for the coupled system
depends sensitively on the disorder and the distribution of
electron phonon coupling
strengths, and may therefore be used to critically compare several
popular
models for the electron phonon coupling constants for these materials.

\section{Disorder and Electronic States in $M_3C_{60}$}

	In this section we review the tight binding description of conduction
electronic states in the doped fullerides.  Since this construction has
been presented in detail elsewhere \cite{GelLu1,Yildirim},
our presentation will be brief.

	The conduction band of the alkali doped fullerides is derived from an
antibonding molecular orbital of the isolated $C_{60}$ with $t_{1u}$
symmetry.  This is the LUMO of the neutral molecule; it is three fold
degenerate, and transforms like a vector under spatial rotations. The
band
structure of the doped and undoped orientationally ordered phases has
been studied by first principles methods \cite{LDA1,LDA2,LDA3}. For both
cases it is found that the
conduction bands can be described reasonably well by a model restricted
to hopping of a three component field $\Psi_i$ = $(c_{i \mu},  \mu =
x,y,z)$  between nearest neighbor fullerene sites:
\begin{equation}
H_{el} = \sum_{i,j} \sum_{\mu \nu} T_{\mu \nu}(i j)
 c^{\dagger}_{i\mu} c_{j \nu} \label{Hel}
\end{equation}
where $c^{\dagger}_{i\mu}$ creates an electron at site $i$ in the $\mu$th
orbital.
The matrix elements in T depend on the orientation of the bond between
sites $i$ and $j$, on the orbital polarizations of the two sites connected in
the hop,
and on the molecular orientations of the terminal sites. In accordance with
experiment \cite{Disordexp} we will restrict
our
attention to the situation where the terminal sites can adopt either of
the two
high symmetry settings with the two fold axes aligned along the (100)
crystal directions,  which we label the ``A" and  ``B" orientations.  The
choice of setting can be indexed by an Ising spin variable,
$s_i$ defined on each
site and which takes the value $s_i$ = 1(-1) for the A(B) orientation.
The
intermolecular hopping amplitude then depends on the spin variables on
neighboring sites in the manner,
\begin{eqnarray}
T(ij) & = & T^{(0)}(\tau_{ij}) + T^{(1)}(\tau_{ij}) s_i
 + T^{(2)}(\tau_{ij}) s_j \\ \nonumber
  & & + T^{(3)}(\tau_{ij}) s_i s_j
\end{eqnarray}
So that the hopping amplitude can be developed from the four fundamental
hopping matrices $T^{(n)}$.  For a hop along a reference bond
$\vec{\tau_{ij}} = (a/2)(\hat{i} + \hat{j}) \equiv \vec{\tau_o}$ the
matrix T take the form

\begin{eqnarray}
  {T^{(0)}(\vec{\tau_o})  =  t \left [ \begin{array}{ccc}
                   A & B & 0 \\
                   B & A & 0 \\
                   0 & 0 & C \end{array} \right ]}  &
  {T^{(1)}(\vec{\tau_o}) = t \left [ \begin{array}{ccc}
                   X & Y & 0 \\
                   -Y & -X & 0 \\
                   0 & 0 & 0 \end{array} \right ]} \nonumber \\
  {T^{(3)}(\vec{\tau_o}) = t \left [ \begin{array}{ccc}
                   X & -Y & 0 \\
                   Y & -X & 0 \\
                   0 & 0 & 0 \end{array} \right ]} &
  {T^{(4)}(\vec{\tau_o})  =  t \left [ \begin{array}{ccc}
                   D & E & 0 \\
                   E & D & 0 \\
                   0 & 0 & F \end{array} \right ]}
\label{HopAmp} \end{eqnarray}
where the three basis orbitals are assigned to the x, y, and z
orbital polarizations, respectively.  The hopping amplitudes for any
other
bond direction can be obtained by a rotation of the T's listed in
equation 3.
The values of the parameters A, B, C, D, E and F , deduced from the work
of Gelfand and Lu, are listed in table 1. We
should note that our convention for assigning the orbital polarizations
is
fixed in the crystal frame and therefore does not change when the
molecular orientation is rotated.  This is in fact different from the
convention
originally adopted by Gelfand and Lu, and the reader is directed to the
Appendix A of Ref. \cite{Yildirim} for a discussion of this point.

	Below we will investigate the effects of orientational disorder in the
supercell
approximation in which we construct a periodically repeated unit cell
containing a large (typically 32) molecules with the orientations
randomly assigned to the A and B settings.
 The electronic Hamiltonian for this model is developed  from the
fundamental hopping matrices listed in equation \ref{HopAmp}. The
artificial periodic boundary conditions allow us to analyze this problem
in the Bloch orbital basis :
\begin{equation}
    c^{\dagger}_{k, i \mu} = \frac{1}{\sqrt{N}} \sum_{I } e^{ik.R_I}
c^{\dagger}_{I, i \mu}
\end{equation}
in a solid with N supercells, indexed by the supercell translation
vectors $R_I$,
and where i  and $\mu$  index  the basis site and orbital polarization
respectively.  In this basis the Hamiltonian for a supercell containing M
sites
is a 3M x 3M matrix H(k) , for which the eigenvectors $C^{\dagger}_{n k}$
are expanded in the Bloch orbital basis:
\begin{equation}
    C^{\dagger}_{n k} =  \sum_{i \mu}   A_{n, i \mu} (k) c^{\dagger}_{k,
i \mu}
\end{equation}

Several authors \cite{GelLu1,GelLu2,MeleErwin1} have applied this
model to study the conduction electron
states in orientationally ordered and disordered structural models.  The
fluctuations in intersite terms in $H_{el}$
 due to the disorder in this
problem are on the order of the Fermi energy,
and so one can correctly anticipate that the effects of disorder will be
quite
significant. Explicit calculations demonstrate that the full conduction
bandwidth in the disordered
phase is only slighly smaller than the bandwidth obtained for the
orientationally ordered structure and in addition the density of states
at the Fermi energy is not significantly different from the value for the
ordered structure. However, the electrons are strongly
scattered by
the orientational fluctuations, and the calculated conduction band
spectrum
for the disordered alloy is smooth and relatively structureless.
A closer inspection of the underlying conduction electron states in the
disordered orientational alloy reveals that  electron propagation near
the
Fermi level can be well described in terms of an effective Bloch
wavevector
k, and a mean free path of order 20 A \cite{MeleErwin1}.
There is no evidence in the
theory for localization of the one electron states in this model except
in a narrow range of
energy  near the band extrema which is irrelvant for electronic processes
near the Fermi energy. This is in accord with experimental measurements
which show a metallic conductivity at low temperature in the normal
phase \cite{ResisT},
that is the measured or extrapolated conductivity at low temperature
comfortably exceeds the minimum metallic conductivities expected at these
electron densities.  Thus theory and experiment provide a consistent
overall
picture that the electronic structure near the Fermi energy for the
disordered alloy is that of a ``dirty"
metal with propagating states at the Fermi energy and a relatively short
residual mean free path.

\section{Intramolecular Phonons}
	In this section we present the model for the ``bare" unrenormalized
phonons of the doped solid.  The 176 modes of a single fullerene
molecule can be indexed using the ten irreducible representations of
the icosahedral group $I_h$ \cite{Group1,Group2}.
However, the conduction states are
derived from the $t_{1u}$ molecular states, and therefore the
conduction charge in this manifold may couple only to the lattice
modes of $a_g$ and $h_g$ symmetry.  There are two such fully
symmetric modes, they are the molecular breathing mode and
pentagonal pinching mode of the icosahedral molecule.  There are eight
families of $h_g$ modes, and these describe the various possible
quadrupolar
deformations of the $C_{60}$ molecule.  These intramolecular modes
span an energy range from 200 $cm^{-1}$ to 1600 $cm^{-1}$.

	In our model, the
bare phonon Hamiltonian will be represented by the Einstein
spectrum:
\begin{equation}
H_{ph} =  \sum_{i} \sum_{\alpha m} \omega_{\alpha m}
 ( b^{\dagger}_{ \alpha m}(i) b_{ \alpha m }(i) + \frac{1}{2} )
\label{Hph}
\end{equation}
where $ b^{\dagger}_{ \alpha m }(i)$ creates the m-th mode of
symmetry $\alpha$ on the i-th site. We shall use a convention where
$\alpha=0$(with $m=1,2$) denotes the $a_g$ modes and $\alpha=1,5$(with
$m=1,8$) label the $h_g$ quintuplets. The degeneracy of each
$h_g$ quintuplet
implies that for $\alpha=1,5$, $\omega_{\alpha m} = \omega_{m}$.
 These even parity modes are experimentally accessible by Raman
scattering, and measurements of the Raman frequencies for $M_x C_{60}$
with x = 0, 3 and 6 have now been reported from several groups
\cite{Raman1,Raman2,Raman3}. The
striking result of the experiments is that the Raman lines in $M_3
C_{60}$
are shifted  and broadened relative to the undoped solid. On the
other hand in $M_6C_{60}$,
which is an insulator, although the frequencies are further shifted,
the Raman
lines are again sharp.  It is suggested that the source of the shift and
broadening in the $M_3 C_{60}$ phase is primarily the coupling of the
intramolecular modes of equation 6 to the conduction charge.  For the
$M_6C_{60}$ structure which is a band  insulator, the charge gap is of order
0.5
eV, so that the phonon self energy is again purely real, leading to a
shift of the frequency but no damping.

	There are several important contributions to the frequency shifts and
linewidths of the Raman active vibrations of the doped system.  Most
notably, the intramolecular interatomic potential is intrinsically
anharmonic.  Thus addition of charge, even to an isolated molecule, will
lead
to a relaxation of the bond lengths, and to a shift of the bare
frequency.
Here ``bare" implies that the phenomenon is not a solid state, but rather
a
molecular effect.  Note that this effect is not  accompanied by
electronic
induced broadening which is a solid state effect.  The frequency
renormalization in the $M_6 C_{60}$ insulating phase is due to just such
an
``intramolecular" effect.

The situation in the interesting $M_3 C_{60}$ phases is complicated due
to
the confluence of these ``intramolecular" and ``intermolecular" solid
state
processes.  The added conduction charge induces an intramolecular
relaxation of the bond lengths and frequency shifts of the type observed
for the $M_6 C_{60}$
phases \cite{Dresselhaus,Faulhaber,Us1}.  There are, in addition,
 self energy contributions due to the ``one
loop" density fluctuations of the added conduction electrons.  (These are
clearly abent for the terminal phase with x=0 which has no conduction
charge, and for the x=6 phase which has a saturated band). These terms
are nevetheless the most interesting since they communicate information
about the solid state structure, and particularly about the intersite
orientational correlations,  into our effective Hamiltonian for the
lattice
degrees of freedom.  These terms are clearly essential for what follows
as they induce a mixing, frequency shift and damping of the coupled
intramolecular modes.

	For simplicity we shall assume that the two types of phonon
renormalizations, intermolecular (due to the one loop corrections from
the
conduction charge) and intramolecular (due to the anharmonic terms in the
bare Hamiltonian and to higer order terms in the electron phonon
coupling)
are simply additive.  Thus the bare Hamiltonian of equation \ref{Hph}
 will be
assumed to already contain the intramolecular effects, and we will
explicitly
calculate the intermolecular contributions absent from that model.  To be
precise, Faulhaber et. al. \cite{Faulhaber} have reported theoretical
results for the lattice
vibrations of the isolated $C_{60} ^{3-}$ trianion, which properly
contain the
intramolecular contributions only, and we adopt their results for the
frequencies $\omega_{\alpha n}$ appearing in equation \ref{Hph}. For
convenience, these frequencies are reproduced in Table 2.  We then
proceed to explicitly evaluate the important  ``solid state" one loop
electronic
contributions to our effective Hamiltonian in the doped phases.

\section{Electron Phonon Interaction}
The lattice modes of equation \ref{Hph} are coupled linearly to the conduction
electrons via an on site deformation potential \cite{Varma,Schluter2}:
\begin{equation}
     H_{el-ph} = \sum_{i} \sum_{\lambda} \sum_{\mu \nu}
          \Gamma_{\mu \nu}^{\lambda} c^{\dagger}_{\mu i} c_{\nu i}
          Q_{\lambda}(i)
\label{EPI}
\end{equation}
where
\begin{equation}
  Q_{\lambda}(i) =  b_{\lambda}(i) +  b^{\dagger}_{\lambda}(i)
\end{equation}
is the displacement corresponding to the $\lambda-th$ mode on site i.
Here $\Gamma_{\mu \nu}^{\lambda} $ is a matrix acting in the
electronic $t_{1u}$ subspace.  The symmetry of this matrix is fixed
solely by the symmetry of the coupled mode $\lambda$ while the
strength of the interaction depends on the particular mode of interest.
If we write $\lambda$ = $\alpha m$ for the m-th mode of symmetry type
$\alpha$ we have that
\begin{equation}
 \Gamma_{\mu \nu}^{\alpha n} = g_{\alpha m} \Delta_{\mu
\nu}^{\alpha}
\label{Gamma}
\end{equation}
where $\alpha$ = 0 denotes a fully symmetric $a_g$ mode  and
$\alpha$ = (1,..,5) denote the five basis functions in the $h_g$
representation of $I_h$.
Anticipating our passage to a crystal with
cubic symmetry we will break this manifold into the invariant $e_g$
and $t_{2g}$ subspaces and write:

\begin{eqnarray}
a_{g} : &
  {\Delta^0 =  \left ( \begin{array}{ccc}
    \frac{1}{\sqrt{3}} & 0 & 0 \\
    0 & \frac{1}{\sqrt{3}} & 0 \\
    0 & 0 &  \frac{1}{\sqrt{3}}\end{array} \right )} \nonumber \\
t_{2g}: &
   {\Delta^1 =  \left ( \begin{array}{ccc}
    0 & \frac{1}{\sqrt{2}} & 0 \\
    \frac{1}{\sqrt{2}} & 0 & 0 \\
     0 & 0 & 0 \end{array} \right )
  \Delta^2 =  \left ( \begin{array}{ccc}
     0 & 0 & 0 \\
     0 & 0 & \frac{1}{\sqrt{2}}  \\
     0 & \frac{1}{\sqrt{2}} & 0  \end{array} \right )} \nonumber \\
  & {\Delta^3 = \left ( \begin{array}{ccc}
    0 & 0 & \frac{1}{\sqrt{2}} \\
    0 & 0 & 0 \\
   \frac{1}{\sqrt{2}} & 0 & 0  \end{array} \right )} \nonumber \\
e_g: & {\Delta^4 =  \left ( \begin{array}{ccc}
         \frac{1}{\sqrt{6}} & 0 & 0 \\
         0  & \frac{1}{\sqrt{6}} & 0 \\
         0 &  0 &  -\frac{2}{\sqrt{6}}  \end{array} \right )
 \Delta^5 = \left ( \begin{array}{ccc}
    \frac{1}{\sqrt{2}} & 0 & 0 \\
      0 & - \frac{1}{\sqrt{2}} & 0 \\
      0 & 0 & 0 \end{array} \right )}
\end{eqnarray}
We have defined these matrices so that they are orthonormal with
$\mbox{tr} \Delta^{\alpha} \Delta^{\beta} = \delta_{\alpha \beta}$.
As expected, the form
of the $\Delta$ matrices requires that the symmetric modes couple
linearly to the total on ball conduction charge density and  the $h_g$
modes are linearly coupled to  the electric quadrupole density
fluctuations.

	The $g_{\alpha m}$'s appearing in equation 9 are the deformation
potentials are obtained by studying the linear variation of the
intramolecular Hamiltonian with respect to the normal coordinates of
the vibrational mode of interest.

\begin{equation}
  g_{\alpha m} \Delta^{\alpha}_{\mu \nu} = \frac{ \hbar}{ M
\omega_{\alpha m} }
 \langle \mu | \frac{ \partial H_{mol} }{ \partial Q_{\alpha m}}| \nu
\rangle
\label{gcoup}
\end{equation}
where $H_{mol}$ is the $\it {intramolecular}$ electronic Hamiltonian
and M is the mass of a carbon atom.  A microscopic evaluation of
these coupling coefficients requires a model for the dependence of
$H_{mol}$ on intramolecular displacements, as well as a microscopic
assignment of the various intramolecular phonon frequencies and
eigenvectors.  There have been several evaluations of the $g_{\alpha
n}$'s reported in the literature \cite{Varma,Schluter1,Faulhaber}
 which treat $H_{mol}$ and the bare vibrational
modes at various levels
of sophistication.   The coefficients $g_{\alpha n}$ reported in these
studies are displayed in table 3. Note that for $h_g$ modes
($\alpha=1,5$), $g_{\alpha m}=g_m$. We also note that the data tabulated here
differ from those reported in several of the original papers owing to
different normalization conditions on the $\Delta$ matrices adopted
by the various authors.
These various models may be classified
 according to two groups: those that localize
the strongest interactions to the two highest frequency $h_g$ modes and
those that distribute the coupling more uniformly over the phonon
spectrum.   The two highest frequency $h_g$ modes contain double bond
stretching character, and therefore might be expected to couple
strongly to the added conduction charge which occupies an
antibonding state.  The coupling to the lower
frequency modes is a more subtle matter, deriving from the
perturbation to $H_{mol}$ due to bond angle fluctuations.

	The expression for $H_{el-ph}$ is valid for an isolated molecule.
Information about the solid state properties is then contained in the
electronic amplitudes $\langle c^{\dagger}_{\mu i} c_{\nu i} \rangle$
appearing in this expression.  Using the notation in Section II for a
periodically repeated supercell, the linearized interaction reads
\begin{equation}
H_{el-ph} =
 \sum_I \sum_{ \lambda i} \sum_{\mu \nu}
           \Gamma_{\mu \nu}^{\lambda} c^{\dagger}_{ I i \mu} c_{ I i \nu}
          Q_{\lambda i}(I)
\end{equation}
Transforming to the basis of Bloch eigenstates $C^{\dagger}_{n k}$
given by the expression,
\begin{equation}
    C^{\dagger}_{n k} =  \sum_{I i \mu}   A_{n, i \mu} (k) e^{i k.R_{I}}
c^{\dagger}_{I,  i \mu}
\end{equation}
we obtain,
\begin{equation}
H_{el-ph} =
    \sum_{k,n,k',n'} \sum_{\lambda i}
       \tilde{\Gamma}^{\lambda i}_{n n'}(k,k')
C^{\dagger}_{ n i}(k)C_{n' i}(k')Q_{\lambda i}(k-k')
\end{equation}
where,
\begin{equation}
  Q_{\lambda i}(q) =  b_{ \lambda i}(-q) +  b^{\dagger}_{ \lambda i}(q)
\end{equation}
and ,
\begin{equation}
  \tilde{\Gamma}^{\lambda i}_{n,n'}(k,k') =
\sum_{ \mu \nu} \Gamma^{\lambda}_{\mu \nu}
 A^{\star}_{n i \mu} (k') A_{n' i \nu}(k)
\end{equation}
is the electron-phonon coupling evaluated in the Bloch basis.
The phonons relevant to our discussion of the infrared
absorption spectra are those with q=0 and hence we will set k = k' in
the above expressions and abbreviate the diagonal components
$\tilde{\Gamma}^{\lambda i}_{n,n'}(k,k)$ by
$\tilde{\Gamma}^{\lambda i}_{n,n'}(k)$ and
$Q_{\lambda i}(0)$ by $Q_{\lambda i}$.

\section{Phonon Self Energies}

	In this section we calculate the one loop corrections to the dynamical
matrix for the phonons in the x=3 phase.
The calculations are carried out first for a
reference orientationally ordered model, and then for a model containing
quenched disorder.  We focus on the selection rules governing the
electron phonon coupling for the ordered case,
 and the breakdown of these rules in
the orientationally disordered phase. We also compute the
renormalized phonon Green's function which includes the
 effects of the one loop corrections
and which will be used to calculate the current current correlation
function for
the coupled electron phonon system in Section VI(b).

	We will study the phonon Green's function
\begin{equation}
 D(\lambda i,\lambda' i';\omega) =
- i \int_0^{\infty} dt e^{-i \omega t}
\langle T [ Q_{\lambda i}, Q^{\dagger}_{\lambda' i'} ] \rangle
\end{equation}
As in the previous section, $\lambda$ indexes the a particular
symmetrized on ball displacement field, and i indexes a site in the unit
cell. We specialize the discussion to the q=0 part of the Green's
function, and  therefore enforce periodic boundary conditions on the
displacement field $Q^{\dagger}_{\lambda i} + Q_{\lambda i}$.  In the
absence of the electron phonon interaction  the unrenormalized
Green's function is diagonal in both the site and phonon ``band" indices
$\lambda$.

\begin{equation}
D_o(\lambda i, \lambda i';\omega) = D_o(\lambda,\omega)
\delta_{\lambda \lambda'} \delta_{i i'}
\end{equation}
where
\begin{equation}
D_o(\lambda,\omega) =\frac{2 \omega_{\lambda} }
   {\omega_{\lambda}^2 - (\omega - i \epsilon_{\lambda})^2 }
\end{equation}
This Green's function is assumed to contain all the relevant
intramolecular renormalizations, which are then parameterized by the
bare frequencies $\omega_{\lambda}$ and the linewidths
$\epsilon_{\lambda}$.  For the former we use the theoretical
data of Ref. \cite{Faulhaber}, while for the latter we use
the experimentally observed linewidths for undoped $C_{60}$
reported in \cite{Raman3}. These parameters are listed in Table 3.

	The linear electron phonon interaction in equation \ref{EPI}
is represented by the elementary vertex shown in Figure 1(a).
In the Bloch orbital basis, this
matrix element is diagonal in the site index but in general not in the
orbital indices $\mu$. This scattering amplitude provides a second
order
contribution to the phonon self energy, where the
relevant  term describes  a process whereby an electron hole pair,
created on a common site $i$ by the vertex shown in Figure 1(a)
propagate and ultimately recombine on a site $i'$ in the solid.  The
vibrational degrees of freedom on different sites are therefore
coupled indirectly through the density fluctuations of the conduction
electrons.  This process, represented diagramatically in Figure 1(b)
, is evaluated to give the
self energy contribution
\begin{eqnarray}
   \Pi(\lambda i, \lambda' i';\omega) &  = \sum_{k} \sum_{n \neq n'}
 & \frac{ \tilde{\Gamma}^{\lambda i}_{n n'}(k)
\tilde {\Gamma}^{\lambda' i'}_{n' n}(k)}
     {E_{nk}-E_{n'k} - \omega - i \delta} \nonumber \\
 &  &(f(E_{nk}) -f(E_{n'k}))
\label{Pi}
\end{eqnarray}
For this molecular solid in which the "bare" phonons are well
described by Einstein oscillators with negligible direct intersite
coupling the indirect coupling in equation \ref{Pi} determines the
dispersion of the vibrational bands of the doped phase.  Similarly, all
the information about the orientational structure of the solid phase
enters
our theory through the spatial variation of the one electronic states
which appear in the polarization contribution in equation \ref{Pi}.

	The separable form of the deformation potential in equation
\ref{Gamma} allows us to write the self-energy term in
equation \ref{Pi} in the factorized form:
 \begin{equation}
   \Pi(\lambda i, \lambda' i';\omega) =
g_{\alpha m} \tilde{\Pi}(\alpha i, \beta i';\omega) g_{\beta m'}
\label{redPi}
\end{equation}
where $\alpha$ labels the phonon symmetry, and n is a phonon band
index.

	For the orientationally ordered phase we can drop
the site index since we have only one molecule per unit cell. In addition
since the icosahedral symmetry is still maintained, the reduced
polarization term $\tilde{\Pi}$ is diagonal in the symmetry labels
$\alpha$, so that
\begin{equation}
 \tilde{\Pi} (\alpha,\beta) = \delta_{\alpha \beta}
 \tilde{\Pi}(\alpha)
\label{OrdPi1}
\end{equation}
Furthermore, since we have chosen the coupling matrices ($\Delta$)
to be irreducible under cubic symmetry we have the additional
simplification,
\begin{equation}
\tilde{\Pi}(0) \neq
\tilde{\Pi}(1) = \tilde{\Pi} (2) = \tilde{\Pi} (3) \neq \tilde{\Pi} (4) =
\tilde{\Pi} (5)
\label{OrdPi2}
\end{equation}
 Ofcourse, the full $\Pi$ in equation \ref{redPi} still
has off-diagonal terms in the phonon band indices. Thus, in the
orientationally ordered solid, the electron
phonon interaction induces coupling only between phonons of the
same symmetry in different bands.

	However, in the orientationally disordered phases, off diagonal
terms can and do appear even in $\tilde{\Pi}$ and they
induce a significant mixing not only between phonons with different
symmetries but also between phonons on different sites. In addition,
there is also a significant variation in the
the diagonal components of $\tilde{\Pi}$ from site to site as we see
below.

	We note that in equation \ref{Pi} we have omitted the long wavelength
contributions from the ``intra-band" terms with $n = n'$.  In fact, for
coupling to the (optical)Einstein modes relevant to our model,  in which
$\omega \rightarrow$ constant as q $\rightarrow$ 0, these contributions
vanish.  This is in striking contrast to the situation for coupling to
acoustic modes, in which an additional {\em intraband} contribution
also exists. Allen \cite{Allen} has derived an approximate expression
for Im($\Pi$) for general wavevector $q$ which is of the form,
\begin{equation}
 \mbox{Im}(\Pi(q)) = \omega \sum_k | g(k,k+q) |^2 \delta(E_k) \delta(E_{k+q})
\end{equation}
The $q=0$ limit of this expression is indeed non-zero and results
in a frequency shift($\Delta \omega_n$) and linewidth($\gamma_n$),
\begin{equation}
 \Delta \omega_n =   N(0) \frac{g_{n}^2}{ d_n};
  \gamma_n = \pi \omega_n N(0)^2 \frac{g_n^2}{d_n}
\end{equation}
where $d_n$ is the degeneracy of the $n$th mode and $N(0)$ is the
density of states at the fermi level.
 Although this
result has been previously applied to the doped fullerides
\cite{Varma,Schluter3}, it is strictly
applicable  only in the limit that the Debye frequency for the relevant
acoustic phonons is much smaller than the fermi energy,
$E_F$, for the conduction charge. These conditions are violated for the
fullerides where we have optical phonons with frequencies
comparable to the fermi energy.
In our treatment we will therefore consider  only the self
energy terms appearing in equation 22.

	Once the self energy contributions are calculated using
equation \ref{Pi} the
renormalized phonon Green's function is obtained by solving the
Dyson equation
\begin{equation}
 D = D_o + D_o \Pi D
\label{Dyson}
\end{equation}
For the orientationally ordered solid, the factorization in
equation (\ref {redPi})
allows us to reduce the matrix equation (\ref{Dyson}) to a scalar
problem in each invariant subspace.
For the
disordered solid in which mixing between the various symmetry
branches and sites occurs one has to solve the full matrix equation.
However,
the mixing between  the $h_g$ and $a_g$ manifolds is expected to be
small and we solve the equation in each subspace.
Due to
the factorisation in (\ref{redPi}) the largest matrix which needs to
be inverted for the $h_g$ phonons is only
5M x 5M(where M is the number of sites).

	For the orientationally ordered phase, the reduced polarization terms
for the fully symmetric space $\alpha$ = 0 vanish identically.
Thus a long wavelength excitation
in the fully symmetric subspace provides a constant shift to the
electronic potential, but cannot scatter electrons between conduction
states.  The remaining nonzero contributions in the irreducible $e_g$
and $t_{2g}$ subspaces are displayed as functions of the phonon
frequency in Figure 2.
The imaginary part of the self energy
(shown in the top panel) introduces a damping of the molecular modes, the
real
part (lower panel) describes the frequency shift due to the screening.
We see that the overall shape of Im $\tilde{\Pi}$ is similar in the two
subspaces, peaking near 2000 $cm^{-1}$.  We find that the $ t_{2g}$
vibrations are coupled slighly more strongly, particularly at higher
frequency.  The most important consequence of this is that the real
part of the self energy at low frequency ( below 1000 $cm^{-1}$ ) is
enhanced in the $ t_{2g}$ channel relative to the $e_g$ channel.
This implies a ``crystal field " splitting of the $\ell $=2
intramolecular modes with the $t_{2g}$ frequency systematically
 suppressed relative to the
$e_g$ frequency.  Solving equation (25) we find that this splitting is of
order $30 cm^{-1}$ for the second $h_g$ mode( $h_g(2)$ in Table. 2) whose
renormalised frequency is 372 $cm^{-1}$.
Finally, we note that the higher frequency
Raman active modes overlap the region in which Im $\tilde{\Pi}$ is
large.  This leads to a relatively large damping of the higher frequency
Raman active vibrations.  However, as we see below this situation is
complicated somewhat by the orientational disorder of the host.

	For the orientationally disordered phase the
symmetry selection rules,(equations (\ref{OrdPi1} and \ref{OrdPi2})
 no longer pertain and there are a large number of independent components
of the reduced self-energy,$\tilde{\Pi}$.
 In Figure 3 we display a representative sample corresponding to
the diagonal components of $\tilde{\Pi}$ corresponding to the same
symmetry branch but different sites(i.e. $\tilde{\Pi}(i \alpha=5,i
\beta=5)$ for various values of site $i$).
It can be seen that there is significant variation in $\tilde{\Pi}$
on different sites. This spread is expected to lead to an additional
inhomogenous broadening of the phonon linewidths.
This shape of these
spectra can be understood simply.  If we model the conduction band
by a rectangular density of states and assume a contact interaction
between the electrons and phonons which is independent of
frequency and wavevector, the predicted Im $\tilde{\Pi} (\omega)$ is
obtained from a simple convolution of the filled and empty sectors of the
one electronic
spectrum.  This yields a density fluctuation spectrum which is
symmetric with a triangular shape, as indeed we find from our
numerical computations.

	The analogous self-energy terms in the
$a_g$ subspace are shown in Figure 4.  The shape is again
"triangular" and the overall scale is similar to that found for
the
$h_g$ subspace. Note however, that the self energy contributions are
quite different
owing to the variation of the amplitudes $g_{\alpha}$. It is
nonetheless significant that these contributions are
$\it{nonzero}$ for the disordered phase.  This reflects the fact that the
long
wavelength phonons in the folded Brillouin zone of the supercell are
coupled to intracell charge fluctuations, and can thus be  screened
and damped by their coupling conduction electrons.  The simplest
example of this nonzero coupling in the symmetric subspace is the
branch which is backfolded from q = $ (\frac {2 \pi} {n a}, 0, 0)$ in the
extended zone, where n is the supercell period.  This couples to the x-
polarized density fluctuation  of the conduction charge in the
supercell.

	We note that there are two intramolecular modes of $a_g$ symmetry:
the breathing mode near 200 $cm^{-1}$ and the pentagonal pinch
near 1500 $cm^{-1}$.  These occur respectively well below, and near,
the peak of Im $\tilde{\Pi} (\omega)$ shown in Figure 4.  Of course
the phonon self energy depends on both the magnitude of $\tilde{\Pi}
(\omega)$ and the coupling strength $g$. The coupling constants
for the $a_g$ modes were computed in Ref. \cite{Schluter1}( see
Table 3) where it was found that the coupling to the breathing
mode was very small while the coupling to the pentagonal pinch mode
is quite substantial. The combination of this large coupling strength with
our calculated $\tilde{\Pi}$ for the $a_g$ symmetry channel would
predict a substantial broadening of the higher frequency $a_g$ mode.
On the other hand experimentally \cite{Raman3}
the increase in the linewidth of
this phonon is small which seems to indicate that the actual
coupling should also be small.

\section{Infrared Conductivity}

	In this section we use the renormalized phonon Green's function
developed in the previous section to calculate the frequency
dependent conductivity of the coupled electron phonon system.  We
begin by considering the purely electronic contributions to the current
current correlation function for the ordered and disordered phases.
We then turn to the indirect contributions to the conductivity mediated
by coupling through the dressed phonons.

\subsection{Electronic Contributions}

	The real part of the frequency dependent conductivity is obtained
within linear response theory as the imaginary part of the fourier
transform of the current current correlation function.  Since the
wavelength of the probe is long on the scale of atomic dimensions it
suffices to study the q = 0 limit of these response functions.   In our
tight binding formalism, the paramagnetic term in the current operator
is given by

\begin{equation}
   \vec{j}   =
       i e \sum_{i j} \sum_{\mu \nu} \vec{\tau_{ij}}T_{\mu \nu}(ij)
                 c^{\dagger}_{i \mu} c_{j \nu}
\end{equation}
where T is the intermolecular hopping matrix along the
$\vec{\tau_{ij}}$-th bond.  Thus the paramagnetic current-current
correlation function is calculated from the Kubo formula
\begin{equation}
  \chi_{a b} (\omega) = i  \int_0 ^{\infty} dt e^{-i \omega t}
            \langle [ j_a (t),j_b (0) ] \rangle
\end{equation}
where the angular brackets denote a thermodynamic average, and a,
b are Cartesian indices.  The frequency dependent conductivity,  which is
the
linearized current in response to an external electric field is therefore
\begin{equation}
  \sigma_{ab}(\omega) = \frac{ Im(\chi_{a b})}{\omega}
\end{equation}
For the noninteracting electronic Hamiltonian introduced in Section II,
this correlation function can be readily calculated from the single
particle eigenstates to yield

\begin{eqnarray}
  (\sigma_{el})_{ab}(\omega) = &
 \frac{1}{N} \sum_{k} \sum_{n  n'}
   \tilde{j}^a_{ n n' }(k) \tilde{j}^b_{n' n}(k) \\ \nonumber
     &  \delta( E_{nk}-E_{n'k} - \omega )
       ( \frac{f(E_{nk}) -f(E_{n'k})}{E_{nk} -E_{n'k}})
\label{sigE}
\end{eqnarray}
where f(E) is the Fermi-Dirac distribution, and $\tilde{j}^a_{ n n' }(k)$
is the matrix element of the  a-th component of the current operator in
(26) between single particle eigenstates n and n'.   In this treatment
which neglects the diamagnetic contributions to j, the long wavelength
low frequency Drude response needs to be evaluated with some care.
One may proceed by studying the intraband contributions to j at finite
wavevector q, and extract the q $\rightarrow$ 0 limit:

\begin{equation}
  (\sigma^{intra}_{ab} (\omega))  =  \sum_{k n}
\tilde{j}^a_{ n n }(k) \tilde{j}^b_{n n}(k)
 \frac {\partial f_{kn}}{\partial E_{nk}} \delta(\omega)
  = D \delta(\omega)
\end{equation}
so that the weight of the Drude pole is obtained as a Fermi surface
average

\begin{equation}
  D =  \sum_{k n} \tilde{j}^a_{ n n }(k) \tilde{j}^b_{n n}(k)
\delta(E_{nk})
\end{equation}
Rather than compute the Fermi surface average required in (30) our
strategy will be to develop the sum rule for the total conductivity

\begin{equation}
    \frac{2}{\pi} \int_{0}^{\infty} d \omega (\sigma_e)_{ab}(\omega)) =
       e^2  \langle K_{ab} \rangle
\end{equation}
where, in the tight binding formalism  $\langle K_{ab} \rangle$ is the
ground state expectation value of the operator:

\begin{equation}
    K_{a b} =  [ R_a,j_b ] =
         \sum_{i j} \sum_{\mu \nu} (\tau_{i j})_a (\tau_{i j})_b
      T_{\mu \nu}(i j) c^{\dagger}_{i\mu} c_{j \nu}
\label{sumrule}
\end{equation}

Equation \ref{sumrule} is the generalized long wavelength
 f-sum rule \cite{SumRule} within our
tight binding model.  For the conducting phase, the interband
paramagnetic part of the response function in \ref{sigE} does not exhaust this
sum rule, so that the residual weight provides the spectral weight in
the Drude pole.  For the free electron gas,  the interband terms vanish
identically, and then full spectral weight is then located in the pole at
zero frequency. Our strategy is therefore to evaluate the sum rule,
and the interband response function directly, and to extract the Drude
piece by a comparision of their respective integrated spectral weights.

	This method can be applied to  both the orientationally
 ordered and
disordered phases. The results for the frequency dependent
conductivity are displayed in Figure 5,
 and the corresponding
partitioning of the interband and "intraband" spectral weights are
given in Table 4. For the orientationally ordered phase we observe
a broad midinfrared continuum, extending from zero frequency to
4000  $cm^{-1}$peaking near 1600 $cm^{-1}$. This peak can be
roughly identified with a peak in the joint density of states do to the
two
primary peaks in the single particle density of states for the ordered
phase.  Approximately 40 $\%$ of the total spectra weight is located in
the Drude pole. We also remark that the interband response function
extends to zero frequency, where it vanishes linearly with $\omega$.
This is due to an interesting symmetry in the band structure for the
orientationally ordered doped phase: the Fermi energy crosses a
doubly degenerate band along the (111) directions in reciprocal
space.  This degeneracy is lifted as on shifts away from this symmetry
direction, yielding a continuum of low frequency interband excitations
which extends exactly to zero frequency.
We remark that this degeneracy has some interesting consequences
for the theory of superconductivity on such a Fermi surface which we
have discussed in a previous paper \cite{MeleErwin2}.

	Our results for a supercell containing  quenched orientational
disorder( also see \cite{GelLu2}) are shown in the lower panel in Figure 5.
The supercell for
this calculation contains 32 sites. Here we see that the interband
peak is broadened and shifts to lower frequency where it overlaps the
overdamped
remnant of the Drude pole.  The composite spectrum thus exhibits an
overdamped midinfrared response function with a distinctly non-Drude
lineshape.  We remark that due to the artificial periodic boundary
conditions in our supercell model, this spectrum retains a very small
but nonzero spectral weight in a true Drude pole at zero frequency.
This weight is nevertheless very small, and accounts for only 2 $\%$ of
the total
spectral weight in this disordered model. This nonzero residual weight
is nevertheless an artifact of the
periodic boundary conditions, and will vanish as the size of the
disordered cell is increased.

	The results of Figure 5 demonstrate that merohedral disorder has a
very strong effect on the dynamics of the conduction electrons, a
point which has been emphasized in several previous studies
\cite{GelLu1,GelLu2,MeleErwin1}.  The
predicted low frequency low temperature conductivity is of order 3000
S/cm.  This is rather small, but significantly larger than the
experimental low
temperature value of 1500 S/cm \cite{Degiorgi}.
 We believe that the discrepancy
arises, at least in part, due to the effects of direct
 electron electron interactions which
have been omitted from our treatment.

\subsection {Lattice Contributions}

	Here we discuss corrections to the direct current current correlation
function  due to the coupling of the conduction charge to the
intramolecular phonons.  We are not concerned with the odd
parity intramolecular vibrations which may be directly coupled to a
long wavelength dipole active probe. Rather we will consider the even
parity
intramolecular modes which are indirectly coupled through the response
conduction electrons.  The appearance of these modes in the infrared
is a very interesting phenomenon which, as we noted previously,  can not
occur for the
orientationally ordered structure, nor indeed any for structure which
preserves the inversion symmetry.  Thus the appearance of these
features in experimentally measured spectra is a fundamental
signature of the microscopic breaking of inversion symmetry by
orientational
disorder. As we see below, these spectra are also quite sensitive to
the details of the electron phonon coupling model, so that
experimental measurements  of this effect might be used as a critical
test of various microscopic models.
	The relevant microscopic process which indirectly couples the even
parity intramolecular modes to light is illustrated diagramatically in
Figure 6.  As with the construction of the ``direct" current current
correlation function, the external field is directly coupled to density
fluctuations of the conduction charge. These may in turn be coupled
to the intramolecular modes through the vertex of Figure 1.  The
composite polarization term is one which thus correlates the current
operator with the vibrational deformation potential  and will be referred
to
as the ``phonon charge." Thus the dressed phonon Green's function
is indirectly coupled to the external field through this effective
phonon charge, contributing to the full current current
correlation function of the coupled electron phonon system.   The
microscopic theory of this process  has been studied for low dimensional
CDW systems \cite{Rice,MJRice} and for the conducting polymers \cite{salts}.
In addition, it has been recently employed to explain the dopant induced
infrared activity of the $t_{1u}$ vibrations of the $C_{60}$ ions
 in the $A_xC_{60}$ compounds \cite{CR}.
	As we noted above, the process represented in Figure 6 must vanish
identically in an inversion symmetric structure. To explicity
demonstrate
this, it is conventient to examine the nonlocal conductivity for the
coupled system in the position representation:

\begin{equation}
    \chi_{ph}(\omega) =  \sum_{ \lambda \lambda'}
   V(\lambda i ; \omega + i \delta)
D(\lambda i, \lambda' j; \omega + i \delta)
   V(\lambda' j;\omega - i \delta)
\end{equation}
where D is the renormalized phonon Green's function , $V(\lambda i
)$ is the electronically induced phonon charge for the $\lambda$ -th
phonon on the i-th site.  $V(\lambda i )$ is given by

\begin{equation}
   V(\lambda i; \omega) = \sum_{i',i''}
 \int dE \mbox{Tr}[  j_{i' i''} G_{i'' i}(E)
 \Upsilon^{\lambda i} G_{i i'}(E+\omega) ]
\end{equation}
where G is a 3x3 matrix one particle Green's function in the space of
orbital polarizations.  In equation (35)  the subscripts i are site
indices
and the trace is taken over the orbital indices. The operator
$\Upsilon^{\lambda i}$ describes the intramolecular scattering of a
conduction electron due to its linear coupling to the $\lambda$-th
mode on site i:

\begin{equation}
   \Upsilon_{\mu \nu}^{\lambda i} = \Gamma^{\lambda}
       c^{\dagger}_{\mu i} c_{\nu i}
\end{equation}
Since the isolated fullerence molecule is inversion symmetric, it is
clear that $\Upsilon^{\lambda i}$  is invariant under the inversion
operation, P.

\begin{equation}
 P^{-1} \Upsilon^{\alpha i} P = \Upsilon^{\alpha i}
\end{equation}
Similarly, the current operator j appearing in equation should
transform as a vector, which is odd under P.

\begin{equation}
 P^{-1} \vec{j} P = - \vec{j}
\end{equation}
We note that on the lattice, the operator j involves the ``off diagonal"
elements of the electronic hopping operator T and thus equation (38)
pertains only in the presence of inversion symmetry.   Finally, for the
inversion symmetric structure the one particle Green's function, and
the two point density response function derived from it is even under
parity.  Thus the phonon charge, which involves a trace over a
product of these three terms is odd under parity, and therefore
vanishes in the q $\rightarrow$ 0 limit.   For the disordered system,
equation (36) is still true.  However, the effective current operator in
(26)
is no longer odd under P, and furthermore, the two point density
response function does not commute with P.   In this case the effective
phonon charge for an intramolecular "even parity" phonon is
nonvanishing.

	We now turn to an evaluation of the of indirect contribution to the
conductivity given in equation (34).   Our disordered system is modelled
by a supercell, with periodic boundary conditions on the molecular
orientations.  In such a model
\begin{equation}
    \chi_{ph}(\omega) =
\sum_{ \lambda i, \lambda' i'}
   V(\lambda i ; \omega + i \delta)
D(\lambda i, \lambda' i'; \omega + i \delta)
   V(\lambda i';\omega - i \delta)
\label{sigph}
\end{equation}
where now $i$ and $i'$ are site indices, and the phonon charge is explicitly:
\begin{eqnarray}
   V(\lambda i;\omega + i\delta) = &  \frac{1}{N} \sum_{k} \sum_{n n'}
 &\frac{ \tilde{j}_{ n n' }(k) \tilde{\Gamma}^{\lambda i}_{n' n}(k) }
      {E_{nk}-E_{n'k} - \omega - i \delta} \nonumber \\
&  &  (f(E_{nk}) -f(E_{n'k}))
\label{phcharge}
\end{eqnarray}
Note that in this model, crystal momentum is conserved modulo a
reciprocal lattice vector of the supercell, and equations(\ref{sigph})
and (\ref{phcharge}) therefore
represent the appropriate q=0 limits of these response functions.

	The effective phonon charge in equation (\ref{phcharge})
is evaluated directly by
numerical integration on a mesh of $12^{3}$ wavevectors in the
folded Brillouin zone reciprocal to a supercell containing 32 molecular
sites. The calculated  contributions to Re $\sigma_{ph}$ due to
coupling to the $h_g$ modes are displayed in Figure 7, where we
have collected results for three different models for the electron
phonon interaction.  These are obtained for each of the coupling
amplitudes $g_{\alpha}$ listed in Table 3. The contributions of
equations (\ref{sigph}) and (\ref{phcharge})
 to the conductivity are additive and therefore each of
these spectra
should be superposed on the direct contribution to the conductivity
shown in Figure 5(b). We see that for all the models, the coupling to
the intramolecular ``quadrupolar" modes produces a substantial
modulation of the otherwise smooth featureless midinfrared conductivity.
These
modulations are correlated with the frequencies of the renormalized
$h_g$ phonons, although it is found that the lineshapes are typically
relatively broad
and quite complex. The overall scale of these
modulations is of order 100 S/cm for all the models studied.
	Interestingly, we observe that the detailed structure in the modulated
midinfrared conductivity is
different for the various models.  As we remarked earlier, the electron
phonon models can be
broadly grouped into two categories.  In the first of these, the electron
phonon
coupling is dominated by the coupling to the two highest lying $h_g$
branches of the spectrum.  The model due to Varma et. al. is a well known
example from this group.  These two highest lying modes are dominated by
double bond stretching character on the fullerene, which are
expected to couple strongly to the "antibonding" $t_{1u}$ electronic
state.   In the second group of models, the coupling is distributed more
uniformly among both the high frequency and lower frequency $h_g$
vibrations.  The models derived by Schluter et. al. and by Faulhaber
et. al. belong to this second category.  These models require
substantial coupling to the bond angle fluctuations and the conduction
electrons.  We find that in these latter models, but not in the former,
the indirect
contributions to the conductivtiy shows two well defined peaks in the
range between 300 and 500 $cm^{-1}$.  In this region, the imaginary
part of the phonon self energy is relatively small, and therefore the
midfrequency intramolecular modes, if coupled to the conduction
charge, show up as sharp features in the spectra.     By contrast, for
the
former class of electron phonon models, with negligible coupling to
the midfrequency modes, the sharp low frequency features are absent and
the modulation spectra are ``rugged" over a
rather wide energy range, extending to the highest frequency modes
near1700 cm-1.

	The process illustrated in Figure 6, also leads to an indirect
contribution to the
conductivity from the high frequency, fully symmetric $a_g$ phonon
branch.
This resulting contribution to the conductivity is displayed in Figure
8
calculated using the theoretically calculated value from Ref.\cite{Schluter1}
Here
the antisymmetric Fano type lineshape which is the signature of the
resonance of this local mode with the pair continuum is clearly
evident. The scale of this modulation is of order 100 S/cm, and its width
is due to the relatively large coupling constant $g_{\alpha}$ for the second
$a_g$ mode listed in Table 3.  Experimentally, this mode is observed as
a much narrower resonance, so that the microscopic  prediction for the coupling
strength for this mode appears to be overestimated.

	Modulations of the midinfrared conductivity
in $K_3 C_{60}$ have recently been
reported in experiments by Kuzmany et. al. \cite{Kuz1,Kuz2}
 In Figure (9) we
reproduce from this work, the real part of the full midinfrared
conductivity, and
the indirect phonon contribution, which is obtained by subtracting a
smooth
electronic background from the experimental measurements.  We note that
the experimental data in this Figure is taken at  T = 300K and the scale
of the
conductivity over this range of frequency is known to depend on
temperature.  With this caveat, the modulations of Re $\sigma$ due to
coupling through the phonons is clearly evident in the experimental
spectra.
The two very sharp features near 600 cm and 1400 cm are odd parity
modes, which are ``directly" dipole active and thus not described by the
indirect
phonon charge model derived above.  The remainder of the modulations in
the
conductivity are on a scale of
typically 25-50 S/cm, and an additional sharp feature due to coupling to
the high
frequency symmetric $a_g$ mode is clearly evident.    These
modulations provide microscopic spectroscopic evidence for the local
breakdown of
inversion symmetry due to  the orientational disorder in these samples.
We also note
that the experiment does not show any sharp phonon structure in the
midfrequency range.  Thus we believe that the type I electron phonon
coupling models provide a more faithful representation of the
experimental
situation in these fullerides than the type II models.

\section{Summary}

	In this paper we have focused on the interference between
the intramolecular
vibrational modes and the low lying single particle excitations measured in the
midinfrared conductivity. Recent experimental data do provide some supporting
evidence for this coupling in the doped phases.  However there are
three aspects of the
measured spectra which remain relatively poorly described by this theory,
and to which
we should call attention. First, the overall scale of the conductivity
 is incorrectly
predicted in the theory. In fact the calculations overestimate the
low frequency limit  of
Re $\sigma ( \omega)$ by nearly a factor of two. This is a serious
difficulty in that the
predicted value of $\sigma (\omega \rightarrow 0)$ is not rescaled by a
simple rescaling
of the conduction bandwidth. We believe that the failure to predict
the correct scale of
the measured conductivity is related to a deeper problem, and is likely
associated with
the neglect of direct electron electron interactions in the model.
Second, careful
experiments at very low frequency show a rather sharp minimum in the frequency
dependent conductivity near $\omega \approx  100 cm^{-1}$.  This is thought to
separate a low frequency Drude like peak, from a rather broad  midinfrared
continuum.  This sharp low frequency minimum is absent from the theory
presented
above, even in the presence of coupling to the phonons.  It is possible that
this originates
in some strong couping to intermolecular vibrations neglected in  our
treatment.
Finally, our discussion does not address the very interesting temperature
dependence of
the resistivity, which increases roughly linearly with temperature over
a very broad
temperature range, extending to 700 K.  At the highest temperatures studied,
one may
infer from these data that the effective mean free path is smaller
than the radius of the
$C_{60}$, a state of affairs which appears difficult to reconcile
with the theoretical
picture given above. Thus, while we believe that the phonon structure in the
midinfared conductivity can be adequately explained by the theory presented
here, there
are a number of issues to be resolved in the experimental low frequency
spectra.

\section{Acknowledgments}

This work was supported by the Department of Energy under Grant
91 ER 45118 and by the NSF under Grant 91 20668.  HYC was supported by the
Korean Science and Engineering Foundation (KOSEF) through the center for
theoretical physics and through grant 931-0200-003-2.

\newpage
{\bf Table 1} Hopping Amplitudes for the electronic Hamiltonian(Equation
\ref{Hel} obtained from Ref. \cite{GelLu1}. These coefficients are
multiplied by a scale factor,$t=0.014$ meV for doped fullerides.
\vspace{.10in}
\begin{center}
\begin{tabular}{|c|c|c|c|c|c|c|c|} \hline
A & B & C & D & E & F & X & Y \\ \hline
0.01 & 0.38 & -2.29 & 2.09 & -2.36 & 0.38 & -0.63 & -0.49 \\ \hline
\end{tabular}
\end{center}

{\bf Table 2.} Frequencies and line-widths for the bare
phonons( in $cm^{-1}$).
The frequencies are as calculated in Ref.\cite{Faulhaber} and
the linewidths are chosen to agree with experiment(\cite{Raman3}).
\vspace{.25in}
\begin{center}
\begin{tabular}{|c|c|c|} \hline
mode     & Frequency & Linewidth   \\ \hline
$h_g(1)$ &  244   & 4.2 \\
$h_g(2)$ &  412  & 5.5 \\
$h_g(3)$ &  560  & 7.5 \\
$h_g(4)$ &  750  & 9.0 \\
$h_g(5)$ &  1073  & 7.0 \\
$h_g(6)$ &  1202  & 7.6 \\
$h_g(7)$ &  1395 & 7.5 \\
$h_g(8)$ &  1508 & 9.5 \\
$a_g(1)$ &  507   &  2.5  \\
$a_g(2)$ &  1470   &  1.5    \\ \hline
\end{tabular}
\end{center}

{\bf Table 3.} Coupling constants,$g_{\alpha n}$, as defined in Equation
\ref{gcoup}, for the $h_g$ and $a_g$ modes(in meV)
\vspace{.25in}
\begin{center}
\begin{tabular}{|c|c|c|c|c|} \hline
mode     & Ref. \cite{Varma} & Ref. \cite{Schluter1}
& Ref. \cite{Faulhaber}  \\ \hline
$h_g(1)$ & 9.4  & 11.6       &   6.0          \\
$h_g(2)$ & 7.1  & 13.5       &  21.1         \\
$h_g(3)$ & 8.9  & 27.6       &   7.6         \\
$h_g(4)$ & 0.0  & 20.5       &  28.7         \\
$h_g(5)$ & 29.0  & 0.0       &  13.4         \\
$h_g(6)$ & 0.0  & 0.0       &   18.0         \\
$h_g(7)$ & 77.8 & 26.5       &  47.8          \\
$h_g(8)$ & 46.0 & 44.0       &  37.8         \\ \hline
$a_g(1)$ & -    & 0.0       &    -           \\
$a_g(2)$ & -    & 52.0       &    -          \\ \hline
\end{tabular}
\end{center}

{\bf Table 4} Spectral weights in units of $ t e^2/ \hbar a$ (where
$a$ is the lattice constant)
corresponding to the interband
and total electronic conductivities.

\vspace{.10in}

\begin{center}
\begin{tabular}{|c|c|c|} \hline
 Model    &   Interband weight & Total weight \\ \hline

Ordered Crystal &   15.2    &   25.0  \\
Disordered Crystal &  26.6 &  27.1\\ \hline
\end{tabular}

\end{center}

\newpage
\begin{center}
{ \bf FIGURE CAPTIONS }
\end{center}

{\bf FIG.1.} (a) The electron-phonon intereaction vertex in the
Bloch orbital basis.(b) The one loop contribution to the phonon
self energy,$\Pi$. The electron and phonon propagators are
represented by a solid and wavy line respectively.

{\bf FIG.2.} Imaginary(upper panel) and
Real(lower panel) parts of the reduced
phonon self-energy( $\tilde{\Pi(\alpha)}$) in the
 orientationally
ordered solid corresponding to the
 $e_g$( $\alpha=1,2$) and
$t_{2g}$( $\alpha=3,4,5$) symmetry channels.

{\bf FIG.3.} Imaginary(upper panel) and Real(lower panel) parts of
the diagonal components of the reduced phonon
self energy($\tilde{\Pi}(i \alpha,i \alpha)$) for $h_g$ phonons
in the disordered solid
corresponding to $\alpha=5$ and a representative sample of sites $i$.

{\bf FIG.4.} Imaginary(upper panel) and Real(lower panel) parts of
the diagonal components of the reduced phonon
self energy($\tilde{\Pi}(i \alpha,i \alpha)$) for $a_g$ phonons
in the disordered solid
corresponding to $\alpha=0$ and a representative sample of sites $i$.

{\bf FIG.5.} Real part of the pure electronic conductivity for the
ordered(upper panel) and disordered(lower panel) solids

{\bf FIG.6.} Diagrammatic representation of the process which generates
the phonon contribution to the conductivity. The solid line
denotes the electron propagator,the wavy line is the renormalised
phonon propagator and the dotted line the photon propagator.

{\bf FIG.7.} Results for the contribution of the $h_g$ modes to the
conductivity using various models for the coupling constants
($g_{\alpha n}$) listed in Table 3.

{\bf FIG.8.} Results for the contribution of the high frequency
$a_g$ mode to the
conductivity using the coupling calculated in Ref.\cite{Schluter1}
(see table 3).

{\bf FIG.9.} (a)The experimental mid-infrared conductivity \cite{Kuz1}
with a fit representing the smooth electronic background (b) The contribution
due to the phonons obtained by subtracting the electronic background.

%
%
%
%
%
%
%
%
%
%
%
%
%
%
%
%
%
\end{document}